\documentclass[preprintnumbers, showpacs, floatfix, preprintnumbers, letterpaper, superscriptaddress, nofootinbib]{revtex4}
\usepackage{amsmath}
\usepackage{amssymb}
\usepackage{graphicx}
\newcommand{\LL}{{\cal L}}
\newcommand{\PP}{{\cal P}}
\newcommand{\be}{\begin{equation}}
\newcommand{\ee}{\end{equation}}
\newcommand{\bea}{\begin{eqnarray}}
\newcommand{\eea}{\end{eqnarray}}

\makeatletter

\newcommand{\Rmnum}[1]{\expandafter\@slowromancap\romannumeral #1@}
\makeatother

\usepackage{color}

\begin{document}

%\title{Magnetogenesis in the bouncing scenario}
\title{Magnetogenesis in bouncing cosmology}

\author{Peng Qian}
\email{qianp@itp.ac.cn}
\affiliation{CAS Key Laboratory of Theoretical Physics, Institute of Theoretical Physics,
Chinese Academy of Sciences, P.O. Box 2735, Beijing 100190, China}
\affiliation{School of Astronomy and Space Science, University of Chinese Academy of Sciences,
No.19A Yuquan Road, Beijing 100049, China}

\author{Yi-Fu Cai}
\email{yifucai@ustc.edu.cn}
\affiliation{CAS Key Laboratory for Researches in Galaxies and Cosmology, Department of Astronomy, University of Science and Technology of China, Chinese Academy of Sciences, Hefei, Anhui 230026, China}

\author{Damien A.~Easson}
\email{easson@asu.edu}
\affiliation{Department of Physics \& Beyond Center for Fundamental Concepts
in Science, Arizona State University, Tempe, AZ 85287-1504, USA}

\author{Zong-Kuan Guo}
\email{guozk@itp.ac.cn}
\affiliation{CAS Key Laboratory of Theoretical Physics, Institute of Theoretical Physics,
Chinese Academy of Sciences, P.O. Box 2735, Beijing 100190, China}
\affiliation{School of Astronomy and Space Science, University of Chinese Academy of Sciences,
No.19A Yuquan Road, Beijing 100049, China}

\begin{abstract}
We consider the process of magnetogenesis in the context of nonsingular bounce cosmology. We show that large primordial magnetic fields can be generated during contraction without encountering strong coupling and backreaction issues. The fields may seed large-scale magnetic fields with observationally interesting strengths. This result leads to a theoretical constraint on the relation of the energy scale of the bounce cosmology to the number of effective e-folding of the contracting phase in the case of scale invariance for the power spectrum of primordial magnetic fields. We show that this constraint can be satisfied in a sizable region of the parameter space for the nonsingular bounce cosmology.
\end{abstract}

\maketitle

%%%%%%%%%%%%%%%%%%%%%%%%%%%%%%%%%%%%%%%%%%%%%%%%%%%%%%%%%%%%%%%%%%%%%%%%%%%%%%%%%%%%%%%%%%%%%%%%%%%%%%%%%%%%%%%%%
\section{Introduction}
Astronomical observations indicate the existence of magnetic fields throughout the observable universe--from the length scales of stars up to the scales of galactic clusters. The strength of these fields in galaxies and clusters is on the order of $\mu$ Gauss. Magnetic fields can be produced from the amplifications of initial seeds via some physical processes during structure formation (such as adiabatic compression and turbulent shock flows). Observations suggest the strengths of initial seed fields should be greater than $10^{-13}$ Gauss on a comoving scale larger than 1 Mpc \cite{Campanelli:2013mea, Campanelli:2015jfa}. If the seed magnetic fields originated in the very early universe, before the surface of last scattering, they may effect the temperature and polarization spectra of the cosmic microwave background (CMB) anisotropies since their energy-momentum tensor can source the corresponding  scalar, vector, and tensor cosmological perturbations. Therefore, CMB observations can provide experimental constrains on the properties of primordial magnetic fields \cite{Durrer:1999bk, Kahniashvili:2000vm, Mack:2001gc, Paoletti:2008ck, Paoletti:2010rx, Paoletti:2012bb, Durrer:2013pga}. According to the latest Planck 2015 data, primordial magnetic field strengths are required to be less than a few $10^{-9}$ gauss at the length scale of 1 Mpc~\cite{Ade:2015cva}.

Discovering the underlying mechanism responsible for generating primordial magnetic seed fields is an active area of cosmological research. In the literature, the origin of primordial magnetic fields has been widely discussed, see Refs.~\cite{Grasso:2000wj, Kandus:2010nw, Yamazaki:2012pg} for relevant reviews. Amongst these mechanisms, one possibility is that the seed fields could be produced during cosmic phase transitions such as the electroweak or QCD phase transitions. This approach, however, suffers from a manifest issue that the correlation lengths are too small when compared with observations. Consequently, one may investigate the possibility of producing seed fields at an even earlier time, such as during the inflationary epoch, in which case the process is known as inflationary magnetogenesis. Since the theory of electrodynamics is conformally invariant, and the energy density of the background universe is nearly constant, inflationary magnetogenesis requires the breaking of conformal invariance by introducing couplings of the electromagnetic fields to other sectors such as gravity \cite{Turner:1987bw, Kunze:2009bs, Kunze:2012rq, BeltranJimenez:2010uh, Jimenez:2010hu},  inflaton fields \cite{Ratra:1991bn, Martin:2007ue, Demozzi:2009fu, Ferreira:2013sqa, Campanelli:2015fna,Adshead:2016iae},  pseudo scalar fields \cite{Turner:1987bw, Garretson:1992vt, Finelli:2000sh, Campanelli:2008kh}, or other fields \cite{Gasperini:1995dh, Bamba:2003av, Bamba:2006ga}. However, these models typically produce seed fields that have too small strengths or suffer from other conceptual issues, such as the backreaction problem that the energy density of the generated electromagnetic fields spoil the inflationary background dynamics, the strong coupling problem that the coupling of electromagnetism (EM) may become too large in the primordial stage. Additionally, there exists the curvature perturbation problem wherein the curvature perturbations seeded by the generated electromagnetic fields are too large to be in agreement with the CMB data~\cite{Barnaby:2012tk, Cai:2010uw, Motta:2012rn, Suyama:2012wh, Shiraishi:2013vja, Fujita:2013pgp}. Recently, two of the current authors proposed a generalized model within the inflationary scenario in which the above issues were avoided under a certain fine tuning \cite{Guo:2015awg}.

One possible alternative to the inflationary paradigm, is that our observed universe may have evolved from a previously contracting phase and then experienced a nonsingular bouncing phase at very early times. This is known as a bouncing scenario \cite{Cai:2014bea, Brandenberger:2009jq, Brandenberger:2016vhg, Novello:2008ra, Lehners:2008vx, Battefeld:2014uga}. In particular, it has been shown that dynamical bounce models under certain contracting phases, such as the matter contraction \cite{Brandenberger:2012zb, Wands:1998yp, Finelli:2001sr} or the ekpyrotic phase \cite{Khoury:2001wf, Lehners:2008vx, Li:2013hga}, a nearly scale invariant power spectrum of primordial fluctuations could be achieved in order to explain the present cosmological observations. Various model constructions and related perturbation analyses were extensively performed in the literature, namely, the quintom bounce~\cite{Cai:2007qw, Cai:2007zv, Cai:2008ed}, the Lee-Wick bounce \cite{Cai:2008qw}, the Horava-Lifshitz gravity bounce \cite{Brandenberger:2009yt, Cai:2009in, Gao:2009wn}, the $f(T)$ teleparallel bounce \cite{Cai:2011tc, deHaro:2012zt, Cai:2015emx}, the ghost condensate bounce \cite{Lin:2010pf}, the Galileon bounce \cite{Qiu:2011cy, Easson:2011zy}, the matter-ekpyrotic bounce \cite{Cai:2012va, Cai:2013kja, Cai:2014zga}, the fermionic bounce \cite{Alexander:2014eva, Alexander:2014uaa}, etc.~(see, e.g. \cite{Brandenberger:2010dk, Brandenberger:2012zb} for recent reviews). In general, it was shown that on length scales larger than the time scale of the nonsingular bouncing phase, primordial cosmological perturbations remain almost unchanged through the bounce \cite{Quintin:2015rta, Battarra:2014tga}.
The bouncing phase embedded in inflationary scenario was recently considered to explain both the large hemispherical power asymmetry and the power deficit on large angular scales in cosmic microwave background radiation~\cite{Liu:2013kea,Liu:2013iha} (see~\cite{Piao:2003zm,Piao:2005ag} for the bouncing inflation).

In this paper we investigate the process of magnetogenesis within the frame of the nonsingular bounce cosmology. The magentogenesis in bouncing universe was once investigated in \cite{Sriramkumar:2015yza, Chowdhury:2016aet} where a specific form of the gauge field coupling was considered. Different from the case in inflationary scenario, the bounce cosmology suggests that our universe had a sufficiently long period of contraction and then experience a smooth bouncing phase that connected the contracting phase with the observed expanding one. Correspondingly, the energy density of the background universe cannot be approximated to be a constant as in the inflationary scenario. It is notoriously difficult to construct bouncing models in General Relativity in a flat universe due to the required violation of the null energy condition (NEC). In general, NEC violation in bouncing cosmology is accompanied by a number of stability problems. If the matter Lagrangian is simply a general function of a scalar field and its derivative (as in k-essence) one cannot transition into an NEC violating phase and hence a bounce is not possible~\cite{Vikman:2004dc, Easson:2016klq}. A construction of a bounce which was manifestly free from gradient and ghost instabilities before thermal expansion and which occurs on trajectories of a non-vanishing measure was achieved in~\cite{Easson:2011zy}.  As a specific model for bouncing cosmology we choose \cite{Cai:2012va} to study magnetogenesis; however, our findings are applicable to general bouncing models. In this paper we show that a bouncing universe can realize magnetogenesis in an efficient way due to the following three conditions. Firstly, as previously mentioned, there exists the large scale problem for the generation of primordial magnetic field. This issue is nicely addressed within the bouncing scenario since in the contracting phase the wavelength of the seed fields exit the Hubble horizon. Secondly, the scale factor in the contraction phase is decreasing and the power spectrum of the magnetic field becomes dense instead of being diluted as in inflation. Thirdly, the energy density of the background universe also becomes dense in the contracting phase, and therefore the backreaction problem can be avoided within a reasonable tuning of model parameters.

The paper is organized as follows. In Section~\Rmnum{2} we introduce the basic formulae of the model to realize magnetogenesis in bouncing cosmology. In Section~\Rmnum{3} we briefly review the scenario of bounce cosmology by considering a specific model but keep the background evolution general. We then analyze the generation of primordial seed fields in Section~\Rmnum{4} and discuss constraints from the strong coupling issue. The backreaction issue of primordial magnetic fields is investigated in Section~\Rmnum{5} within the bouncing scenario. We summarize our results in Section~\Rmnum{6}.

%%%%%%%%%%%%%%%%%%%%%%%%%%%%%%%%%%%%%%%%%%%%%%%%%%%%%%%%%%%%%%%%%%%%%%%%%%%%%%%%%%%%%%%%%%%%%%%%%%%%%%%%%%%%%%%%%
%\section{Model}\label{sec:model}

\section{The model of bounce magnetogenesis}

To begin, we consider the following action for scalar field $\phi$ and electromagnetic field $A_\mu$, minimally coupled to Einstein gravity
\begin{equation}
 S=\int d^4x \sqrt{-g}\left[\frac12 R+\LL_{\rm bounce}-\frac{1}{4}f^2(\phi)F^2\right],
\label{eq:action}
\end{equation}
where $\LL_{\rm bounce}$ is the Lagrangian density responsible for the nonsingular bouncing background which will be introduced in detail in Section \Rmnum{3}, and
the field strength $F_{\mu\nu}$, is defined in terms of the $U(1)$ gauge field in the familiar way $F_{\mu\nu} = \partial_\mu A_\nu - \partial_\nu A_\mu$. Unless explicitly stated, we will work with natural units in which $c=\hbar=M_{pl}=1$.

Moreover, the standard Lagrangian of the electromagnetic field can be recovered when the coefficient $f$ is taken to be unity. In order to achieve the amplification of the magnetic fields, one expects the coefficient $f$ to be a function of the background scalar field $\phi$ and hence can evolve along with the cosmic time so that the conformal symmetry is violated under this time evolution.

Varying the above action \eqref{eq:action} with respect to the electromagnetic field yields the corresponding equation of motion as follows,
\begin{equation}
 \nabla_{\mu}\left[-f^2F^{\mu \nu }\right]=0.
\end{equation}
We consider an observationally favored, spatially flat Friedmann-Robertson-Walker metric,
\begin{eqnarray} \label{FRWmetric}
 ds^2=a^2(\eta)(-d\eta^2+d\mathbf{x}^2),
\end{eqnarray}
where $a(\eta)$ is the scale factor and $\eta$ is the conformal time.

We choose the Coulomb gauge: $A_0=0$, $\partial_{i}A^{i}=0$, and expand $A_i$ in the Fourier space as follows,
\begin{eqnarray}
 A_i(\eta, \mathbf{x}) = \sum_{\sigma=1,2}\int\frac{d^3k}{(2\pi)^{3/2}}
  \left(\epsilon_{i,\sigma}(\mathbf{k}) a_{\mathbf{k},\sigma} A_{k}(\eta) e^{i\mathbf{k}\cdot\mathbf{x}}+{\rm h.c.}\right),
\end{eqnarray}
where $\epsilon_{i,\sigma}(\mathbf{k})$ are two orthonormal and transverse polarization vectors. The operators $a_{\mathbf{k},\sigma}$ and $a_{\mathbf{k},\sigma}^{\dagger}$ are the annihilation and creation operators that satisfy the standard commutation relations $[a_{\mathbf{k},\sigma},a_{\mathbf{k}',\sigma'}^{\dagger}] =\delta_{\sigma\sigma'}\delta(\mathbf{k}-\mathbf{k}')$ as well as $[a_{\mathbf{k},\sigma},a_{\mathbf{k}',\sigma'}] = [a_{\mathbf{k},\sigma}^{\dagger},a_{\mathbf{k}',\sigma'}^{\dagger}]=0$.

Within the metric \eqref{FRWmetric}, the Fourier mode $A_k$ with a fixed conformal wave number $k$ satisfies the following equation of motion,
\begin{eqnarray}
 A_k''+\frac{2f'}{f} A_k'+k^2A_k=0~,
\label{eq:me}
\end{eqnarray}
with the normalization condition
\bea
A_{k}(\eta)A_{k}'^{*}(\eta)-A_{k}^{*}(\eta)A_{k}'(\eta)=\frac{i}{f^2}~,
\eea
being satisfied. In terms of a new variable $v_k = fA_k$, the equation of motion \eqref{eq:me} can be reformulated as:
\begin{eqnarray}
v_k''+\left(k^2-\frac{f''}{f}\right)v_k=0.
\label{eq:newme}
\end{eqnarray}
From this equation, one can explicitly read that the evolution of the magnetic field, which is depicted by the gauge field $A_i$, depends on the form of the coupling coefficient $f$ and the background evolution. In the following section we investigate the background of a bouncing universe.

\section{Bounce cosmology}

Nonsingular bounce cosmologies appear in many theoretical frameworks where the gravitational sector is modified, by making use of Null Energy Condition violating matter fields, or in a universe with non-flat spatial geometry (see e.g. \cite{Martin:2003sf, Solomons:2001ef}). However, it is important to note that the BKL instability could appear in the contracting phase since the back-reaction of anisotropies would dominate over the regular dust and radiation densities, unless one finely tunes the initial conditions to be nearly perfectly isotropic. This issue can be avoided if a period of the ekpyrotic phase is introduced before the bounce \cite{Erickson:2003zm} where the unwanted anisotropies are depressed in the presence of an ekpyrotic scalar field.

Recently, it was shown that nonsingular bouncing cosmology with regular matter contraction can be achieved with an era of ekpyrotic contraction by introducing a scalar field with a Horndeski-type, non-standard kinetic term and a negative exponential potential \cite{Cai:2012va}. One could include a regular dust component and then assume the universe began in a state of matter-dominated contraction thus combining the matter bounce with the ekpyrotic scenario. The backreaction of anisotropies is manageable throughout the entire cosmological evolution of the matter-ekpyrotic bounce model, including at the bounce point \cite{Cai:2013vm}, and hence, the BKL instability that arises for a large family of non-singular bounce models, as pointed out in \cite{Xue:2010ux, Xue:2011nw}, is nicely resolved. A concrete realization of the matter-ekpyrotic bounce was constructed in \cite{Cai:2013kja} that involves two matter fields. This effective field theory model of a non-singular bounce can be embedded into a supersymmetric version \cite{Koehn:2013upa}, or into loop quantum cosmology \cite{Cai:2014zga} (also see \cite{WilsonEwing:2012pu, Wilson-Ewing:2013bla, Amoros:2014tha} for different cosmological scenarios).

In the present study, we follow the original bounce model within the context of effective field approach that was proposed in \cite{Cai:2012va} and its cosmological implications to magnetogenesis. The lagrangian can be written as
\begin{eqnarray}
\LL_{\rm bounce}=-K(\phi,X)-G(\phi,X)\square\phi,
\end{eqnarray}
where $K$ and $G$ are functions of the dimensionless scalar field $\phi$ and its canonical kinetic term $X = \partial_{\mu}\phi \partial^{\mu}\phi/2$. The d'Alembertian operator is given by $\square\phi \equiv g^{\mu\nu}\nabla_{\mu}\nabla_{\nu}\phi$.

Following \cite{Cai:2012va} the function $K$ is chosen as
\begin{eqnarray}
 K(\phi,X)=(1-g(\phi))X+\beta X^2-V(\phi),
\end{eqnarray}
and $G$ as
\begin{eqnarray}
G(X)=\gamma X,
\end{eqnarray}
where $\beta$, $\gamma$ are the related model parameters. For a phase of ekpyrotic contraction, the form of the potential $V(\phi)$ is taken to be
\begin{eqnarray}
 V(\phi)=-\frac{2V_{0}}{e^{-\sqrt{\frac{2}{q}}\phi}+e^{b_{V}\sqrt{\frac{2}{q}}\phi}},
\end{eqnarray}
where $V_0$, $q$ ,$b_V$ are model parameters as well. In order to obtain a violation of the Null Energy Condition near the end of the contraction phase, we take $g(\phi)$ to be
\begin{eqnarray}
 g(\phi)=\frac{2g_{0}}{e^{-\sqrt{\frac{2}{p}}\phi}+e^{b_{g}\sqrt{\frac{2}{p}}\phi}},
\end{eqnarray}
where $g_{0}$,$b_{g}$ and $p$ are model parameters whose reasonable range can be found in reference \cite{Cai:2012va}.

The evolution equation of the background dynamics are evaluated explicitly in this model, and are well studied in the literature. We do not repeat this study here, but rather provide a summary of the three stages of evolution most relevant to the process of magnetogenesis:
\begin{itemize}
\item the ekpyrotic contraction,
\item the bouncing phase,
\item the fast-roll phase.
\end{itemize}
Note that, compared to the first phase, the effects of the latter two are rather limited based on the analyses in \cite{Quintin:2015rta}, and hence, we treat them as instantaneous for simplicity. Moreover, we are interested in the contraction phase, during which the wavelength of the fields exits the Hubble horizon since this length scale contracts much faster than the scale factor. It is important to notice that during the same stage the energy density of the scalar field $\phi$ evolves as a perfect fluid with an almost constant equation of state $w$.

During this phase $\phi\ll-1$ so that $g\rightarrow0$ and $\dot{\phi}\ll1$. The Lagrangian for $\phi$ approaches the standard canonical form. There is an attractor solution for $\phi$ which yields an effective equation of state,
\begin{eqnarray}
 w\simeq-1+\frac{2}{3q}.
\end{eqnarray}
If $q$ has a critical value, the equation of state $w$ can be sufficiently large and the energy density of $\phi$ will dominate over matter and radiation during contraction. The background in this case can be described as
\begin{eqnarray}
 a\varpropto(\tilde{\eta}_{e}-\eta)^{\frac{q}{1-q}},
\end{eqnarray}
where the subscript $e$ stands for the end of the contraction phase, $\tilde{\eta}_{e}=\eta_{e}-\frac{q}{(1-q)a_{e}H_{e}}$.
$\eta_{e}$ is the end of the contraction phase and $\eta=0$ is the midpoint of the bounce scenario. The above description is valid until we reach $\eta_{e}$. For simplicity, we define $\tilde{q}=q/(1-q)$ and the scale factor can be expressed as
\begin{eqnarray}
 a(\eta)=a_{e}\left(-\frac{\tilde{q}}{a_{e}H_{e}}\right)^{-\tilde{q}}(\tilde{\eta}_{e}-\eta)^{\tilde{q}},
\label{eq:sf}
\end{eqnarray}
where $a_e$ is the scale factor at the end of the contraction phase.

After the moment $\eta_e$, the universe can evolve through the nonsingular bouncing phase smoothly. During this phase, the evolution of the Hubble parameter $H$ can be approximated as a linear function of the cosmic time $t$ \cite{Cai:2012va, Cai:2013kja}. Accordingly, the scale factor behaves as $a \sim \exp(t^2)$ roughly. In this case, one can read that a nonsingular bounce can occur when the scale factor reaches the minimal value.

In the following section we perform the analysis of the generation of the primordial magnetic fields in the aforementioned cosmological scenario.

\section{Bounce magnetogenesis}

After the bounce we wish to recover classical electromagnetism in order to connected with the observed big bang history smoothly, and therefore we require $f\to 1$ along the background evolution. If in the contraction phase $f$ is much smaller than unity we will face the so-called the strong coupling problem. A natural way to solve such a problem is to require $f^2>1$ during the contraction phase and
%after the bounce it becomes unity.
it becomes unity at the end moment of the contracting phase $\eta_e$, which is also the beginning of the bouncing phase.
This can be done by assuming that, during the contracting phase when $\eta<\eta_e$, $f$ is a power-law function of the scale factor via
\begin{eqnarray}
 f = f_e\left(\frac{a}{a_e}\right)^n,
\label{eq:cf}
\end{eqnarray}
where $f_e\approx{\cal{O}}(1)$. Since in the contraction phase the scale factor decreases, we take $n>0$ so that $f>1$ is fulfilled. During the contraction phase, the Hubble horizon decreases faster than the scale factor so that the modes with larger wavelengths will exited the Hubble horizon earlier.

For short wavelength modes, or $k^2\gg f''/f$, the solution to Eq.~\eqref{eq:newme} is approximated as
\begin{eqnarray}
 v_k(\eta) = \frac{1}{\sqrt{2k}}e^{-ik\eta},
\label{eq:inis}
\end{eqnarray}
while  long wavelength modes ($k^2\ll f''/f$) have the general solution,
\begin{eqnarray}
 v_k(\eta) = C_1 f +C_2 f\int\frac{d\eta}{f^2},
\end{eqnarray}
where $C_1$ and $C_2$ are integration constants that can be fixed by matching this solution to~\eqref{eq:inis}. With the background $a\varpropto(\tilde{\eta}_{e}-\eta)^{\tilde{q}}$ and integrating the second term we  find
\begin{eqnarray}
 v_k(\eta) = C_1 a^n + C_2 a^{-1-n+\frac{1}{q}}.
\end{eqnarray}
If $-1-n+1/q<n$, the second term will dominate over the first in the contraction phase. Using the conjunction condition \begin{eqnarray}
 C_2 a_k^{-1-n+\frac{1}{q}} = \frac{1}{\sqrt{2k}} |e^{-ik \eta_{k}}|,
\end{eqnarray}
where $a_k$ stands for the scale factor when the corresponding wavelength exits the Hubble horizon $a_{k}|H|=k$ (which is also equivalent to $k(\tilde{\eta}_{e}-\eta_{k})=\tilde{q}$) one derives
\begin{eqnarray}
 C_2 = \frac{1}{\sqrt{2k}}a_{e}^{n+1-\frac{1}{q}}\left(-\frac{a_{e}H_{e}}{k}\right)^{\tilde{q}(n+1-\frac{1}{q})}.
\end{eqnarray}
We arrive at the complete form for $v_k$,
\begin{eqnarray}
 v_k = \frac{1}{\sqrt{2k}} \left(-\frac{a_{e}H_{e}}{k}\right)^{\tilde{q}(n+1-\frac{1}{q})} \left(\frac{a}{a_{e}}\right)^{-n-1+\frac{1}{q}}.
\end{eqnarray}
Actually, with the scale factor given by Eq.~\eqref{eq:sf} and the coupling function given by Eq.~\eqref{eq:cf},
the analytical solution to Eq.~\eqref{eq:newme} is a linear combination of Bessel functions~\cite{Ferreira:2013sqa}.
The two integration constants can be fixed by imposing the Bunch-Davies Vacuum condition~\eqref{eq:inis} as $(-k\eta) \to \infty$.
For modes well outside the Hubble radius $(k^2 \ll f''/f)$,
one can get the same solution as (22) in the case of $-1-n+1/q<n$. Therefore, the gauge field is found to be
\begin{eqnarray}
 A_k = v_k/f = \frac{1}{\sqrt{2k}f_{e}} \left(-\frac{a_{e}H_{e}}{k}\right)^{\tilde{q}(n+1-\frac{1}{q})} \left(\frac{a}{a_{e}}\right)^{-2n-1+\frac{1}{q}}.
\end{eqnarray}

Correspondingly, the power spectrum of the primordial magnetic fields is evaluated as
\begin{equation}
 \PP_B(k,\eta) = \sum_{\sigma=1,2} \frac{k^5|A_k(\eta)|^2}{2\pi^2a^4} = \frac{1}{2\pi^{2}f_{e}^{2}} \left(-\frac{a_{e}H_{e}}{k}\right)^{\frac{2q(n+3)-6}{1-q}} \left(\frac{a}{a_{e}}\right)^{-4n-6+\frac{2}{q}}H_{e}^{4},
\end{equation}
with the spectral index $n_B=(6-2(3+n)q)/(1-q)$. Note that the spectrum is scale-invariant for $n=-3+3/q$. At the end of the contraction phase we have
\begin{equation}
 \PP_B(k,a_{e}) = \frac{1}{2\pi^{2}f_{e}^{2}} \left(-\frac{a_{e}H_{e}}{k}\right)^{\frac{2q(n+3)-6}{1-q}}H_{e}^{4}.
\label{eq:ps}
\end{equation}

As argued previously, the effects of the (fast) bounce are limited and hence we treat this phase instantaneously.
Note that the vector field has become the standard electromagnetic field right before the bounce at the moment $\eta_e$. After that,
%After this phase the universe evolves into a standard expanding phase where the gauge field becomes the standard electromagnetic field.
the magnetic fields decay with the square of scale factor. The strength of the magnetic field observed today is given by:
\begin{equation}
 B_{0} = (\PP_B(k,a_{0}))^{1/2} = \frac{1}{2\pi f_{e}} \left(-\frac{a_{e}H_{e}}{k}\right)^{\frac{q(n+3)-3}{1-q}} \frac{a_{e}^{2}}{a_{0}^{2}}H_{e}^{2},
\label{eq:ps}
\end{equation}
where $a_{0}$ is the scale factor today. In the scale-invariant case we have
\begin{equation}
 B_{0} = \frac{1}{\sqrt{2}\pi f_{e}} H_{e}^{2}\frac{a_{e}^{2}}{a_{0}^{2}}.
\label{eq:ps}
\end{equation}
We must determine the amount the scale factor changes from the end of the contraction until today. Using the conservation of entropy we find
\begin{equation}\label{a_e_0}
 \frac{a_e}{a_0} \simeq \frac{T_0}{\rho_{EM}^{1/4}}.
\end{equation}

\begin{figure}[htbp]
\centering
\includegraphics[height=4cm,width=0.4\linewidth]{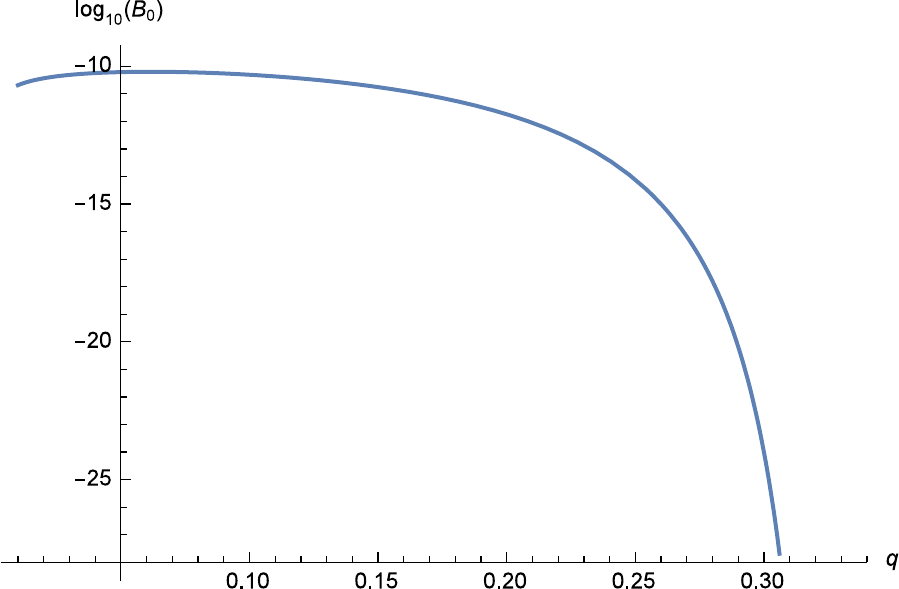}
\caption{Dependence of the upper bound of the magnetic fields $B_0$ on the model parameter $q$ in the bounce model under consideration. Other model parameters have been taken to be the same as those provided in Ref.~\cite{Cai:2013vm}. }
\label{figgg}
\end{figure}

We treat the energy density of the electric field as the energy density of radiation at the end of the contraction phase because it dominates over that of magnetic fields. The power spectrum of the electric fields takes the form of
\begin{equation}
 \PP_E(k,\eta) = \sum_{\sigma=1,2} \frac{k^3|A_k'(\eta)|^2}{2\pi^2a^4} = \frac{(2n+1-\frac{1}{q})^2}{2\pi^{2}f_{e}^{2}} \left(-\frac{a_{e}H_{e}}{k}\right)^{2\tilde{q}(n+1-\frac{1}{q})-2} \left(\frac{a}{a_{e}}\right)^{-4n-4}H_{e}^4.
\label{eq:ps}
\end{equation}
The ratio of $\PP_E$ to $\PP_B$ can be written as
\begin{eqnarray}
 \frac{\PP_{E}}{\PP_{B}} = (2n+1-\frac{1}{q})^2 \left(\frac{a}{a_e}\right)^{2-\frac{2}{q}} \left(\frac{a_{e}H_{e}}{k}\right)^2,
\end{eqnarray}
from which we can see that $\PP_{E}\gg\PP_{B}$ for the long wavelengthes beyond the Hubble horizon. Therefore, we can compute the energy density of the electric part as the total energy density of the electromagnetic fields
\begin{eqnarray}\label{rho_EM}
\begin{aligned}
 \rho_{EM} &= f^2\int^{a_{e} H_{e}}_{a_{i}H_{i}} \PP_E \frac{dk}{k}\\
 &= \frac{(-1+q+2nq)^2(1-q) (a_{e}H_{e})^{2\tilde{q} (n+1-\frac{1}{q})-2}H_{e}^{4}}{2\pi^{2}q^2(4-2(2+n)q)} \left(\frac{a}{a_{e}}\right)^{-2n-4} \Big[ (a_e H_e)^{\frac{4-(4+2n)q}{1-q}} -(a_i H_i)^{\frac{4-(4+2n)q}{1-q}} \Big],
\end{aligned}
\end{eqnarray}
where $a_{i}$ is the scale factor at the moment when the largest scale observed today first exited the Hubble horizon. Since at the end of the contraction phase $a_{i} H_{i} \ll a_{e}H_{e}$ in the case of the scale-invariant power spectrum, the first term on the right-hand side of equations \eqref{rho_EM} can be ignored. Thus, we have
\begin{eqnarray}\label{rho_EM_e}
 \rho_{EM}(a_{e}) \simeq \frac{25(1-q)^2a_{e}^{2}H_{e}^{6}}{2\pi^2q^2}(a_i H_i)^{-2},
\end{eqnarray}
From equations \eqref{rho_EM_e} and \eqref{a_e_0}, we get
\begin{eqnarray}
 H_{e}^{2}\frac{a_{e}^{2}}{a_{0}^{2}}\sim\frac{\sqrt{2}T_{0}^2\pi q}{5(1-q)}\frac{a_{i}H_{i}}{a_{e}H_{e}}.
\end{eqnarray}
The strength of the magnetic fields today is then given by
\begin{eqnarray}
 B_{0} = \frac{T_{0}^2 q}{5f_{e}(1-q)} \frac{a_{i}H_{i}}{a_{e}H_{e}} = \frac{T_{0}^2 q}{5f_{e}(1-q)}\exp(-N),
\label{eq:ps}
\end{eqnarray}
where $N$ is the effective e-folding number of the contraction phase. Following Ref. \cite{Cai:2013vm}, it is limited by
\begin{eqnarray}
 N>\frac{1-q}{2(1-3q)}\ln\left(\frac{1}{\Omega_{pert}}\right),
\end{eqnarray}
where $\Omega_{pert}$ is the amplitude of the observed anisotropy in CMB. This leads to
\begin{eqnarray}
 B_{0} \leq \frac{T_{0}^2 q}{5f_{e}(1-q)} \Omega_{pert}^{\frac{1-q}{2(1-3q)}},
\label{eq:ps}
\end{eqnarray}
Choosing $f_{e}=1$ and $q=0.1$, and using $T_{0} = 1.9\times10^{-32}$ in Planck units and $\Omega_{pert}=10^{-5}$, we have the upper bound of the strength of the magnetic fields today:
\begin{eqnarray}
 B_{0} \leq 10^{-11}\text{Gauss}.
\label{eq:ps}
\end{eqnarray}
Note that, we need to ensure that the contracting phase proceeds for a sufficiently long period as discussed above, and the condition that $-1-n+1/q<n$ has to be satisfied in the scale-invariant case. The combination of these two conditions then yields the upper bound of $B_0$ as a function of the model parameter $q$. In Fig.~\ref{figgg}, we demonstrate the relation between the upper bound of $B_0$ and the parameter $q$ for specific parameter choices.

\section{Constraints from the backreaction}

In the above section we performed the analyses of the magnetogenesis within bounce cosmology and showed that the corresponding results can satisfy the observational constraints which gives rise to an upper bound for the produced magnetic fields. In this section we further check the conditions that the energy density of the electromagnetic fields does not spoil the evolution of the background dynamics, which requires
\begin{eqnarray}
 \rho_{EM}<\rho_{\phi},
\end{eqnarray}
where $\rho_{\phi}$ stands for the energy density of the background\footnote{Note that, if there exist degrees of freedom other than the background field $\phi$, such as the radiation, a second matter field, or even the cosmic anisotropy, the stability of the bouncing phase can be well under control in certain parameter space of the bounce model, as has been comprehensively studied in \cite{Cai:2013vm}. From \cite{Cai:2013vm}, one can see that a nonsingular bounce can be obtained when the total energy density is vanishing. In this case, the background dynamics of the universe is dominated by the pressure of the field $\phi$ instead of energy densities. As a result, the requirement $\rho_{EM}<\rho_{\phi}$ remains valid near the bounce.}. Since $\phi$ dominates over other components at late times, we may compare it with the contribution of the electromagnetic fields directly for simplicity. From \eqref{rho_EM} we find
\begin{eqnarray}
 \rho_{EM} \propto a^{-2n-4}\propto a^{2-\frac{6}{q}} ,
\end{eqnarray}
in the case of scale invariance. Moreover, the energy density of the background scaler field evolves as
\begin{eqnarray}
 \rho_{\phi}\propto a^{-3(w+1)}\propto a^{-\frac{2}{q}}.
\end{eqnarray}

The above two equations imply that the energy density of EM grow even faster than that of $\phi$. We require that at the end of the contraction phase $\phi$ dominates,
that is to say,
\begin{eqnarray}
\rho_{EM}(a_{e}) < H_{e}^{2}M_{pl}^2,
\end{eqnarray}
where we recover the Planck energy scale unit $M_{pl}$ to balance the units on both sides of the above equation.
This gives
\begin{eqnarray}
\frac{25(1-q)^2a_{e}^{2}H_{e}^{4}}{2\pi^2q^2(a_{i}H_{i})^{2}M_{pl}^{2}} < 1,
\end{eqnarray}
which leads to
\begin{eqnarray}\label{constraint_t}
\frac{|H_{e}|}{M_{pl}} < \frac{\sqrt{2}\pi q\exp(-N)}{5(1-q)}.
\end{eqnarray}
As a result, the inequality yields a theoretical constraint on the relation of the effective e-foldings $N$ to the energy scale of $\phi$ at the end of contraction phase.

\begin{figure}[htbp]
\centering
\includegraphics[height=4cm,width=0.4\linewidth]{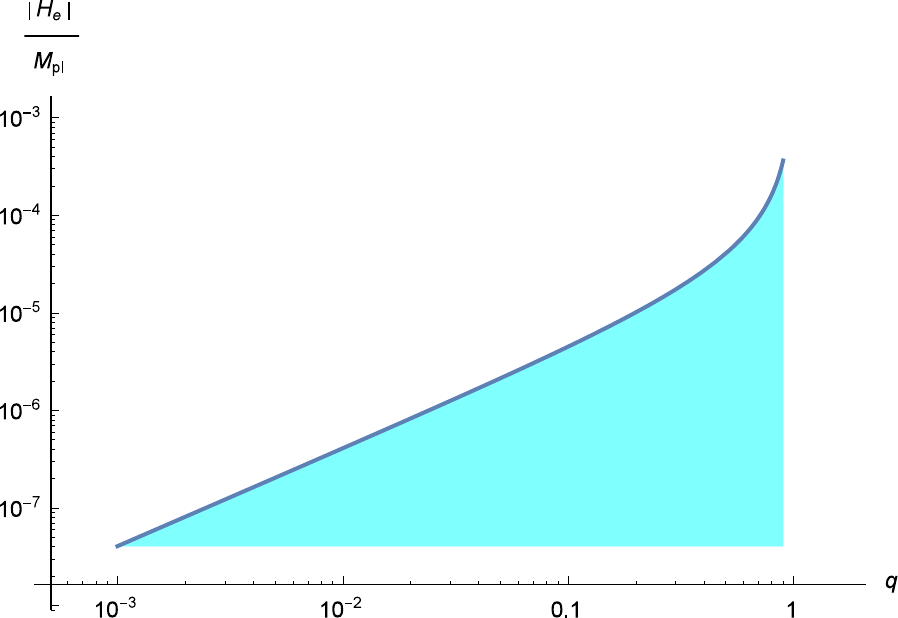}
\caption{The theoretical constraint of the energy scale of the bounce cosmology $|H_e|/M_{pl}$ along with the model parameter $q$. The green shadow area satisfies the theoretical constraint \eqref{constraint_t} and hence bounce models within this regime are well behaved. We take the effective e-folding number to be $N=10$ as an example. }
\label{figgg2}
\end{figure}

In order to have a clearer picture of this theoretical constraint, we numerically depict the parameter space that is allowed for a well behaved bounce model in Fig.~\ref{figgg2}. We find this theoretical constraint is easily satisfied in a sizable region of the parameter space of nonsingular bounce cosmologies.

\section{Conclusion}

In this paper we showed it is possible to produce appreciable primordial magnetic fields within the context of nonsingular bounce cosmology. We adopted the specific model developed in \cite{Cai:2012va} but the results derived in our study are applicable to more general bouncing models as the main analyses only depend on the phase of ekpyrotic contraction. In this model, we found that primordial magnetic field with potentially sizable strength could be produced without troublesome strong coupling and backreaction effects, which plague the process of magnetogenesis in other very early universe paradigms such as inflation. By focusing our interest on the nearly scale-invariance power spectrum of primordial magnetic fields, we derived an explicit observational constraint upon the relation between the magnetic field that may be observed today and the model parameter that is associated with the effective equation of state during the contracting phase. Moreover, by examining the limit of the backreaction, we obtained a theoretical constraint on the relation of the energy scale of the bounce cosmology with respect to the effective e-folding numbers of the contracting phase in the case of scale invariance. We showed that this theoretical constraint can be satisfied in a sizable parameter space of bounce cosmologies.

\begin{acknowledgements}
We thank J. Chen for valuable comments on our manuscript. This work is supported in part by the National Natural Science Foundation of China Grants No.11575272 and No.11335012, in part by the Strategic Priority Research Program of CAS Grant No.XDB23030100 and by the Key Research Program of Frontier Sciences of CAS Grant No.QYZDJ-SSWSYS006.
YFC is supported in part by the Chinese National Youth Thousand Talents Program, by the USTC start-up funding (Grant No. KY2030000049) and by the National Natural Science Foundation of China (Grant No. 11421303).
\end{acknowledgements}

%%%%%%%%%%%%%%%%%%%%%%%%%%%%%%%%%%%%%%%%%%%%%%%%%%%%%%%%%%%%%%%%%%%%%%%%%%%%%%%%%%%%%%%%%%%%%%%%%%%%%%%%%%%%%%%%%


\begin{thebibliography}{99}

%\cite{Campanelli:2013mea}
\bibitem{Campanelli:2013mea}
  L.~Campanelli,
  %``Origin of Cosmic Magnetic Fields,''
  Phys.\ Rev.\ Lett.\  {\bf 111}, no. 6, 061301 (2013)
  [arXiv:1304.6534 [astro-ph.CO]].
  %%CITATION = ARXIV:1304.6534;%%

%\cite{Campanelli:2015jfa}
\bibitem{Campanelli:2015jfa}
  L.~Campanelli,
  %``Lorentz-violating inflationary magnetogenesis,''
  Eur.\ Phys.\ J.\ C {\bf 75}, no. 6, 278 (2015)
  [arXiv:1503.07415 [gr-qc]].
  %%CITATION = ARXIV:1503.07415;%%

%\cite{Durrer:1999bk}
\bibitem{Durrer:1999bk}
  R.~Durrer, P.~G.~Ferreira and T.~Kahniashvili,
  %``Tensor microwave anisotropies from a stochastic magnetic field,''
  Phys.\ Rev.\ D {\bf 61}, 043001 (2000)
  [astro-ph/9911040].
  %%CITATION = ASTRO-PH/9911040;%%

%\cite{Kahniashvili:2000vm}
\bibitem{Kahniashvili:2000vm}
  T.~Kahniashvili, A.~Kosowsky, A.~Mack and R.~Durrer,
  %``Cmb signatures of a primordial magnetic field,''
  AIP Conf.\ Proc.\  {\bf 555}, 451 (2001)
  [astro-ph/0011095].
  %%CITATION = ASTRO-PH/0011095;%%

%\cite{Mack:2001gc}
\bibitem{Mack:2001gc}
  A.~Mack, T.~Kahniashvili and A.~Kosowsky,
  %``Microwave background signatures of a primordial stochastic magnetic field,''
  Phys.\ Rev.\ D {\bf 65}, 123004 (2002)
  [astro-ph/0105504].
  %%CITATION = ASTRO-PH/0105504;%%

%\cite{Paoletti:2008ck}
\bibitem{Paoletti:2008ck}
  D.~Paoletti, F.~Finelli and F.~Paci,
  %``The full contribution of a stochastic background of magnetic fields to CMB anisotropies,''
  Mon.\ Not.\ Roy.\ Astron.\ Soc.\  {\bf 396}, 523 (2009)
  [arXiv:0811.0230 [astro-ph]].
  %%CITATION = ARXIV:0811.0230;%%

%\cite{Paoletti:2010rx}
\bibitem{Paoletti:2010rx}
  D.~Paoletti and F.~Finelli,
  %``CMB Constraints on a Stochastic Background of Primordial Magnetic Fields,''
  Phys.\ Rev.\ D {\bf 83}, 123533 (2011)
  [arXiv:1005.0148 [astro-ph.CO]].
  %%CITATION = ARXIV:1005.0148;%%

%\cite{Paoletti:2012bb}
\bibitem{Paoletti:2012bb}
  D.~Paoletti and F.~Finelli,
  %``Constraints on a Stochastic Background of Primordial Magnetic Fields with WMAP and South Pole Telescope data,''
  Phys.\ Lett.\ B {\bf 726}, 45 (2013)
  [arXiv:1208.2625 [astro-ph.CO]].
  %%CITATION = ARXIV:1208.2625;%%

%\cite{Durrer:2013pga}
\bibitem{Durrer:2013pga}
  R.~Durrer and A.~Neronov,
  %``Cosmological Magnetic Fields: Their Generation, Evolution and Observation,''
  Astron.\ Astrophys.\ Rev.\  {\bf 21}, 62 (2013)
  [arXiv:1303.7121 [astro-ph.CO]].
  %%CITATION = ARXIV:1303.7121;%%

%\cite{Ade:2015cva}
\bibitem{Ade:2015cva}
  P.~A.~R.~Ade {\it et al.} [Planck Collaboration],
  %``Planck 2015 results. XIX. Constraints on primordial magnetic fields,''
  arXiv:1502.01594 [astro-ph.CO].
  %%CITATION = ARXIV:1502.01594;%%

%\cite{Grasso:2000wj}
\bibitem{Grasso:2000wj}
  D.~Grasso and H.~R.~Rubinstein,
  %``Magnetic fields in the early universe,''
  Phys.\ Rept.\  {\bf 348}, 163 (2001)  [astro-ph/0009061].
  %%CITATION = ASTRO-PH/0009061;%%

%\cite{Kandus:2010nw}
\bibitem{Kandus:2010nw}
  A.~Kandus, K.~E.~Kunze and C.~G.~Tsagas,
  %``Primordial magnetogenesis,''
  Phys.\ Rept.\  {\bf 505}, 1 (2011)  [arXiv:1007.3891 [astro-ph.CO]].
  %%CITATION = ARXIV:1007.3891;%%

%\cite{Yamazaki:2012pg}
\bibitem{Yamazaki:2012pg}
  D.~G.~Yamazaki, T.~Kajino, G.~J.~Mathew and K.~Ichiki,
  %``The Search for a Primordial Magnetic Field,''
  Phys.\ Rept.\  {\bf 517}, 141 (2012)  [arXiv:1204.3669 [astro-ph.CO]].
  %%CITATION = ARXIV:1204.3669;%%

%\cite{Turner:1987bw}
\bibitem{Turner:1987bw}
  M.~S.~Turner and L.~M.~Widrow,
  %``Inflation Produced, Large Scale Magnetic Fields,''
  Phys.\ Rev.\ D {\bf 37}, 2743 (1988).
  %%CITATION = PHRVA,D37,2743;%%

%\cite{Kunze:2009bs}
\bibitem{Kunze:2009bs}
  K.~E.~Kunze,
  %``Large scale magnetic fields from gravitationally coupled electrodynamics,''
  Phys.\ Rev.\ D {\bf 81}, 043526 (2010) [arXiv:0911.1101 [astro-ph.CO]].
  %%CITATION = doi:10.1103/PhysRevD.81.043526;%%

%\cite{Kunze:2012rq}
\bibitem{Kunze:2012rq}
  K.~E.~Kunze,
  %``Completing magnetic field generation from gravitationally coupled electrodynamics with the curvaton mechanism,''
  Phys.\ Rev.\ D {\bf 87}, no. 6, 063505 (2013) [arXiv:1210.6899 [astro-ph.CO]].
  %%CITATION = doi:10.1103/PhysRevD.87.063505;%%

%\cite{BeltranJimenez:2010uh}
\bibitem{BeltranJimenez:2010uh}
  J.~Beltran Jimenez and A.~L.~Maroto,
  %``Dark energy, non-minimal couplings and the origin of cosmic magnetic fields,''
  JCAP {\bf 1012}, 025 (2010) [arXiv:1010.4513 [astro-ph.CO]].
  %%CITATION = doi:10.1088/1475-7516/2010/12/025;%%

%\cite{Jimenez:2010hu}
\bibitem{Jimenez:2010hu}
  J.~Beltran Jimenez and A.~L.~Maroto,
  %``Cosmological magnetic fields from inflation in extended electromagnetism,''
  Phys.\ Rev.\ D {\bf 83}, 023514 (2011) [arXiv:1010.3960 [astro-ph.CO]].
  %%CITATION = doi:10.1103/PhysRevD.83.023514;%%

%\cite{Ratra:1991bn}
\bibitem{Ratra:1991bn}
  B.~Ratra,
  %``Cosmological 'seed' magnetic field from inflation,''
  Astrophys.\ J.\  {\bf 391}, L1 (1992).
  %%CITATION = ASJOA,391,L1;%%

%\cite{Martin:2007ue}
\bibitem{Martin:2007ue}
  J.~Martin and J.~Yokoyama,
  %``Generation of Large-Scale Magnetic Fields in Single-Field Inflation,''
  JCAP {\bf 0801}, 025 (2008)  [arXiv:0711.4307 [astro-ph]].
  %%CITATION = ARXIV:0711.4307;%%

%\cite{Demozzi:2009fu}
\bibitem{Demozzi:2009fu}
  V.~Demozzi, V.~Mukhanov and H.~Rubinstein,
  %``Magnetic fields from inflation?,''
  JCAP {\bf 0908}, 025 (2009)  [arXiv:0907.1030 [astro-ph.CO]].
  %%CITATION = ARXIV:0907.1030;%%

%\cite{Ferreira:2013sqa}
\bibitem{Ferreira:2013sqa}
  R.~J.~Z.~Ferreira, R.~K.~Jain and M.~S.~Sloth,
  %``Inflationary magnetogenesis  without the strong coupling problem,''
  JCAP {\bf 1310}, 004 (2013)  [arXiv:1305.7151 [astro-ph.CO]].
  %%CITATION = ARXIV:1305.7151;%%

%\cite{Campanelli:2015fna}
\bibitem{Campanelli:2015fna}
  L.~Campanelli,
  %``Cosmic magnetization out from the vacuum,''
  arXiv:1508.01247 [astro-ph.CO].
  %%CITATION = ARXIV:1508.01247;%%

%\cite{Adshead:2016iae}
\bibitem{Adshead:2016iae}
  P.~Adshead, J.~T.~Giblin, T.~R.~Scully and E.~I.~Sfakianakis,
  %``Magnetogenesis from axion inflation,''
  arXiv:1606.08474 [astro-ph.CO].
  %%CITATION = ARXIV:1606.08474;%%
  %1 citations counted in INSPIRE as of 02 Aug 2016

%\cite{Garretson:1992vt}
\bibitem{Garretson:1992vt}
  W.~D.~Garretson, G.~B.~Field and S.~M.~Carroll,
  %``Primordial magnetic fields from pseudoGoldstone bosons,''
  Phys.\ Rev.\ D {\bf 46}, 5346 (1992)  [hep-ph/9209238].
  %%CITATION = HEP-PH/9209238;

%\cite{Finelli:2000sh}
\bibitem{Finelli:2000sh}
  F.~Finelli and A.~Gruppuso,
  %``Resonant amplification of gauge fields in expanding universe,''
  Phys.\ Lett.\ B {\bf 502}, 216 (2001)  [hep-ph/0001231].
  %%CITATION = HEP-PH/0001231;%%

%\cite{Campanelli:2008kh}
\bibitem{Campanelli:2008kh}
  L.~Campanelli,
  %``Helical Magnetic Fields from Inflation,''
  Int.\ J.\ Mod.\ Phys.\ D {\bf 18}, 1395 (2009)  [arXiv:0805.0575 [astro-ph]].
  %%CITATION = ARXIV:0805.0575;%%

%\cite{Gasperini:1995dh}
\bibitem{Gasperini:1995dh}
  M.~Gasperini, M.~Giovannini and G.~Veneziano,
  %``Primordial magnetic fields from string cosmology,''
  Phys.\ Rev.\ Lett.\  {\bf 75}, 3796 (1995)  [hep-th/9504083].
  %%CITATION = HEP-TH/9504083;%%

%\cite{Bamba:2003av}
\bibitem{Bamba:2003av}
  K.~Bamba and J.~Yokoyama,
  %``Large scale magnetic fields from inflation in dilaton electromagnetism,''
  Phys.\ Rev.\ D {\bf 69}, 043507 (2004)  [astro-ph/0310824].
  %%CITATION = ASTRO-PH/0310824;%%

%\cite{Bamba:2006ga}
\bibitem{Bamba:2006ga}
  K.~Bamba and M.~Sasaki,
  %``Large-scale magnetic fields in the inflationary universe,''
  JCAP {\bf 0702}, 030 (2007)  [astro-ph/0611701].
  %%CITATION = ASTRO-PH/0611701;%%

%\cite{Barnaby:2012tk}
\bibitem{Barnaby:2012tk}
  N.~Barnaby, R.~Namba and M.~Peloso,
  %``Observable non-gaussianity from gauge field production in slow roll inflation, and a challenging connection with magnetogenesis,''
  Phys.\ Rev.\ D {\bf 85}, 123523 (2012)  [arXiv:1202.1469 [astro-ph.CO]].
  %%CITATION = ARXIV:1202.1469;%%

%\cite{Cai:2010uw}
\bibitem{Cai:2010uw}
  R.~G.~Cai, B.~Hu and H.~B.~Zhang,
  %``Acoustic signatures in the Cosmic Microwave Background bispectrum from primordial magnetic fields,''
  JCAP {\bf 1008}, 025 (2010)  [arXiv:1006.2985 [astro-ph.CO]].
  %%CITATION = ARXIV:1006.2985;%%

%\cite{Motta:2012rn}
\bibitem{Motta:2012rn}
  L.~Motta and R.~R.~Caldwell,
  %``Non-Gaussian features of primordial magnetic fields in power-law inflation,''
  Phys.\ Rev.\ D {\bf 85}, 103532 (2012)  [arXiv:1203.1033 [astro-ph.CO]].
  %%CITATION = ARXIV:1203.1033;%%

%\cite{Suyama:2012wh}
\bibitem{Suyama:2012wh}
  T.~Suyama and J.~Yokoyama,
  %``Metric perturbation from inflationary magnetic field and generic bound on inflation models,''
  Phys.\ Rev.\ D {\bf 86}, 023512 (2012)  [arXiv:1204.3976 [astro-ph.CO]].
  %%CITATION = ARXIV:1204.3976;%%

%\cite{Shiraishi:2013vja}
\bibitem{Shiraishi:2013vja}
  M.~Shiraishi, E.~Komatsu, M.~Peloso and N.~Barnaby,
  %``Signatures of anisotropic sources in the squeezed-limit bispectrum of the cosmic microwave background,''
  JCAP {\bf 1305}, 002 (2013) [arXiv:1302.3056 [astro-ph.CO]].
  %%CITATION = doi:10.1088/1475-7516/2013/05/002;%%

%\cite{Fujita:2013pgp}
\bibitem{Fujita:2013pgp}
  T.~Fujita and S.~Yokoyama,
  %``Higher order statistics of curvature perturbations in IFF model and its Planck constraints,''
  JCAP {\bf 1309}, 009 (2013) [arXiv:1306.2992 [astro-ph.CO]].
  %%CITATION = doi:10.1088/1475-7516/2013/09/009;%%

%\cite{Guo:2015awg}
\bibitem{Guo:2015awg}
  P.~Qian and Z.~K.~Guo,
  %``Model of inflationary magnetogenesis,''
  Phys.\ Rev.\ D {\bf 93}, no. 4, 043541 (2016)
  %doi:10.1103/PhysRevD.93.043541
  [arXiv:1512.05050 [astro-ph.CO]].
  %%CITATION = doi:10.1103/PhysRevD.93.043541;%%

%\cite{Cai:2014bea}
\bibitem{Cai:2014bea}
  Y.~F.~Cai,
  %``Exploring Bouncing Cosmologies with Cosmological Surveys,''
  Sci.\ China Phys.\ Mech.\ Astron.\  {\bf 57}, 1414 (2014)
  %doi:10.1007/s11433-014-5512-3
  [arXiv:1405.1369 [hep-th]].
  %%CITATION = doi:10.1007/s11433-014-5512-3;%%

%\cite{Brandenberger:2009jq}
\bibitem{Brandenberger:2009jq}
  R.~H.~Brandenberger,
  %``Alternatives to the inflationary paradigm of structure formation,''
  Int.\ J.\ Mod.\ Phys.\ Conf.\ Ser.\  {\bf 01}, 67 (2011)
  %doi:10.1142/S2010194511000109
  [arXiv:0902.4731 [hep-th]].
  %%CITATION = doi:10.1142/S2010194511000109;%%

%\cite{Brandenberger:2016vhg}
\bibitem{Brandenberger:2016vhg}
  R.~Brandenberger and P.~Peter,
  %``Bouncing Cosmologies: Progress and Problems,''
  arXiv:1603.05834 [hep-th].
  %%CITATION = ARXIV:1603.05834;%%

%\cite{Novello:2008ra}
\bibitem{Novello:2008ra}
  M.~Novello and S.~E.~P.~Bergliaffa,
  %``Bouncing Cosmologies,''
  Phys.\ Rept.\  {\bf 463}, 127 (2008)
  %doi:10.1016/j.physrep.2008.04.006
  [arXiv:0802.1634 [astro-ph]].
  %%CITATION = doi:10.1016/j.physrep.2008.04.006;%%

%\cite{Lehners:2008vx}
\bibitem{Lehners:2008vx}
  J.~L.~Lehners,
  %``Ekpyrotic and Cyclic Cosmology,''
  Phys.\ Rept.\  {\bf 465}, 223 (2008)
  %doi:10.1016/j.physrep.2008.06.001
  [arXiv:0806.1245 [astro-ph]].
  %%CITATION = doi:10.1016/j.physrep.2008.06.001;%%

%\cite{Battefeld:2014uga}
\bibitem{Battefeld:2014uga}
  D.~Battefeld and P.~Peter,
  %``A Critical Review of Classical Bouncing Cosmologies,''
  Phys.\ Rept.\  {\bf 571}, 1 (2015)
  %doi:10.1016/j.physrep.2014.12.004
  [arXiv:1406.2790 [astro-ph.CO]].
  %%CITATION = doi:10.1016/j.physrep.2014.12.004;%%

%\cite{Brandenberger:2012zb}
\bibitem{Brandenberger:2012zb}
  R.~H.~Brandenberger,
  %``The Matter Bounce Alternative to Inflationary Cosmology,''
  arXiv:1206.4196 [astro-ph.CO].
  %%CITATION = ARXIV:1206.4196;%%

%\cite{Wands:1998yp}
\bibitem{Wands:1998yp}
  D.~Wands,
  %``Duality invariance of cosmological perturbation spectra,''
  Phys.\ Rev.\ D {\bf 60}, 023507 (1999)
  %doi:10.1103/PhysRevD.60.023507
  [gr-qc/9809062].
  %%CITATION = doi:10.1103/PhysRevD.60.023507;%%

%\cite{Finelli:2001sr}
\bibitem{Finelli:2001sr}
  F.~Finelli and R.~Brandenberger,
  %``On the generation of a scale invariant spectrum of adiabatic fluctuations in cosmological models with a contracting phase,''
  Phys.\ Rev.\ D {\bf 65}, 103522 (2002)
  %doi:10.1103/PhysRevD.65.103522
  [hep-th/0112249].
  %%CITATION = doi:10.1103/PhysRevD.65.103522;%%

%\cite{Khoury:2001wf}
\bibitem{Khoury:2001wf}
  J.~Khoury, B.~A.~Ovrut, P.~J.~Steinhardt and N.~Turok,
  %``The Ekpyrotic universe: Colliding branes and the origin of the hot big bang,''
  Phys.\ Rev.\ D {\bf 64}, 123522 (2001)
  [hep-th/0103239].
  %%CITATION = HEP-TH/0103239;%%

%\cite{Li:2013hga}
\bibitem{Li:2013hga}
  M.~Li,
  %``Note on the production of scale-invariant entropy perturbation in the Ekpyrotic universe,''
  Phys.\ Lett.\ B {\bf 724}, 192 (2013)
  [arXiv:1306.0191 [hep-th]].
  %%CITATION = ARXIV:1306.0191;%%

%\cite{Cai:2007qw}
\bibitem{Cai:2007qw}
  Y.~F.~Cai, T.~Qiu, Y.~S.~Piao, M.~Li and X.~Zhang,
  %``Bouncing universe with quintom matter,''
  JHEP {\bf 0710}, 071 (2007)
  %doi:10.1088/1126-6708/2007/10/071
  [arXiv:0704.1090 [gr-qc]].
  %%CITATION = doi:10.1088/1126-6708/2007/10/071;%%

%\cite{Cai:2007zv}
\bibitem{Cai:2007zv}
  Y.~F.~Cai, T.~Qiu, R.~Brandenberger, Y.~S.~Piao and X.~Zhang,
  %``On Perturbations of Quintom Bounce,''
  JCAP {\bf 0803}, 013 (2008)
  %doi:10.1088/1475-7516/2008/03/013
  [arXiv:0711.2187 [hep-th]];
  %%CITATION = doi:10.1088/1475-7516/2008/03/013;%%

%\cite{Cai:2008ed}
\bibitem{Cai:2008ed}
  Y.~F.~Cai and X.~Zhang,
  %``Evolution of Metric Perturbations in Quintom Bounce model,''
  JCAP {\bf 0906}, 003 (2009)
  %doi:10.1088/1475-7516/2009/06/003
  [arXiv:0808.2551 [astro-ph]];
  %%CITATION = doi:10.1088/1475-7516/2009/06/003;%%

%\cite{Cai:2008qw}
\bibitem{Cai:2008qw}
  Y.~F.~Cai, T.~t.~Qiu, R.~Brandenberger and X.~m.~Zhang,
  %``A Nonsingular Cosmology with a Scale-Invariant Spectrum of Cosmological Perturbations from Lee-Wick Theory,''
  Phys.\ Rev.\ D {\bf 80}, 023511 (2009)
  %doi:10.1103/PhysRevD.80.023511
  [arXiv:0810.4677 [hep-th]].
  %%CITATION = doi:10.1103/PhysRevD.80.023511;%%

%\cite{Brandenberger:2009yt}
\bibitem{Brandenberger:2009yt}
  R.~Brandenberger,
  %``Matter Bounce in Horava-Lifshitz Cosmology,''
  Phys.\ Rev.\ D {\bf 80}, 043516 (2009)
  %doi:10.1103/PhysRevD.80.043516
  [arXiv:0904.2835 [hep-th]].
  %%CITATION = doi:10.1103/PhysRevD.80.043516;%%

%\cite{Cai:2009in}
\bibitem{Cai:2009in}
  Y.~F.~Cai and E.~N.~Saridakis,
  %``Non-singular cosmology in a model of non-relativistic gravity,''
  JCAP {\bf 0910}, 020 (2009)
  %doi:10.1088/1475-7516/2009/10/020
  [arXiv:0906.1789 [hep-th]].
  %%CITATION = doi:10.1088/1475-7516/2009/10/020;%%

%\cite{Gao:2009wn}
\bibitem{Gao:2009wn}
  X.~Gao, Y.~Wang, W.~Xue and R.~Brandenberger,
  %``Fluctuations in a Horava-Lifshitz Bouncing Cosmology,''
  JCAP {\bf 1002}, 020 (2010)
  %doi:10.1088/1475-7516/2010/02/020
  [arXiv:0911.3196 [hep-th]].
  %%CITATION = doi:10.1088/1475-7516/2010/02/020;%%

%\cite{Cai:2011tc}
\bibitem{Cai:2011tc}
  Y.~F.~Cai, S.~H.~Chen, J.~B.~Dent, S.~Dutta and E.~N.~Saridakis,
  %``Matter Bounce Cosmology with the f(T) Gravity,''
  Class.\ Quant.\ Grav.\  {\bf 28}, 215011 (2011)
  %doi:10.1088/0264-9381/28/21/215011
  [arXiv:1104.4349 [astro-ph.CO]].
  %%CITATION = doi:10.1088/0264-9381/28/21/215011;%%

%\cite{deHaro:2012zt}
\bibitem{deHaro:2012zt}
  J.~de Haro and J.~Amoros,
  %``Nonsingular Models of Universes in Teleparallel Theories,''
  Phys.\ Rev.\ Lett.\  {\bf 110}, no. 7, 071104 (2013)
  %doi:10.1103/PhysRevLett.110.071104
  [arXiv:1211.5336 [gr-qc]].
  %%CITATION = doi:10.1103/PhysRevLett.110.071104;%%

%\cite{Cai:2015emx}
\bibitem{Cai:2015emx}
  Y.~F.~Cai, S.~Capozziello, M.~De Laurentis and E.~N.~Saridakis,
  %``f(T) teleparallel gravity and cosmology,''
  arXiv:1511.07586 [gr-qc].
  %%CITATION = ARXIV:1511.07586;%%

%\cite{Lin:2010pf}
\bibitem{Lin:2010pf}
  C.~Lin, R.~H.~Brandenberger and L.~Perreault Levasseur,
  %``A Matter Bounce By Means of Ghost Condensation,''
  JCAP {\bf 1104}, 019 (2011)
  %doi:10.1088/1475-7516/2011/04/019
  [arXiv:1007.2654 [hep-th]].
  %%CITATION = doi:10.1088/1475-7516/2011/04/019;%%

%\cite{Qiu:2011cy}
\bibitem{Qiu:2011cy}
  T.~Qiu, J.~Evslin, Y.~F.~Cai, M.~Li and X.~Zhang,
  %``Bouncing Galileon Cosmologies,''
  JCAP {\bf 1110}, 036 (2011)
  %doi:10.1088/1475-7516/2011/10/036
  [arXiv:1108.0593 [hep-th]].
  %%CITATION = doi:10.1088/1475-7516/2011/10/036;%%

%\cite{Easson:2011zy}
\bibitem{Easson:2011zy}
  D.~A.~Easson, I.~Sawicki and A.~Vikman,
  %``G-Bounce,''
  JCAP {\bf 1111}, 021 (2011)
  %doi:10.1088/1475-7516/2011/11/021
  [arXiv:1109.1047 [hep-th]].
  %%CITATION = doi:10.1088/1475-7516/2011/11/021;%%

%\cite{Cai:2012va}
\bibitem{Cai:2012va}
  Y.~F.~Cai, D.~A.~Easson and R.~Brandenberger,
  %``Towards a Nonsingular Bouncing Cosmology,''
  JCAP {\bf 1208}, 020 (2012)
  %doi:10.1088/1475-7516/2012/08/020
  [arXiv:1206.2382 [hep-th]].
  %%CITATION = doi:10.1088/1475-7516/2012/08/020;%%

%\cite{Cai:2013kja}
\bibitem{Cai:2013kja}
  Y.~F.~Cai, E.~McDonough, F.~Duplessis and R.~H.~Brandenberger,
  %``Two Field Matter Bounce Cosmology,''
  JCAP {\bf 1310}, 024 (2013)
  %doi:10.1088/1475-7516/2013/10/024
  [arXiv:1305.5259 [hep-th]].
  %%CITATION = doi:10.1088/1475-7516/2013/10/024;%%

%\cite{Cai:2014zga}
\bibitem{Cai:2014zga}
  Y.~F.~Cai and E.~Wilson-Ewing,
  %``Non-singular bounce scenarios in loop quantum cosmology and the effective field description,''
  JCAP {\bf 1403}, 026 (2014)
  %doi:10.1088/1475-7516/2014/03/026
  [arXiv:1402.3009 [gr-qc]].
  %%CITATION = doi:10.1088/1475-7516/2014/03/026;%%

%\cite{Alexander:2014eva}
\bibitem{Alexander:2014eva}
  S.~Alexander, C.~Bambi, A.~Marciano and L.~Modesto,
  %``Fermi-bounce Cosmology and scale invariant power-spectrum,''
  Phys.\ Rev.\ D {\bf 90}, no. 12, 123510 (2014)
  %doi:10.1103/PhysRevD.90.123510
  [arXiv:1402.5880 [gr-qc]].
  %%CITATION = doi:10.1103/PhysRevD.90.123510;%%

%\cite{Alexander:2014uaa}
\bibitem{Alexander:2014uaa}
  S.~Alexander, Y.~F.~Cai and A.~Marciano,
  %``Fermi-bounce cosmology and the fermion curvaton mechanism,''
  Phys.\ Lett.\ B {\bf 745}, 97 (2015)
  %doi:10.1016/j.physletb.2015.04.026
  [arXiv:1406.1456 [gr-qc]].
  %%CITATION = doi:10.1016/j.physletb.2015.04.026;%%

%\cite{Brandenberger:2010dk}
\bibitem{Brandenberger:2010dk}
  R.~H.~Brandenberger,
  %``Cosmology of the Very Early Universe,''
  AIP Conf.\ Proc.\  {\bf 1268}, 3 (2010)
  %doi:10.1063/1.3483879
  [arXiv:1003.1745 [hep-th]].
  %%CITATION = doi:10.1063/1.3483879;%%

%\cite{Quintin:2015rta}
\bibitem{Quintin:2015rta}
  J.~Quintin, Z.~Sherkatghanad, Y.~F.~Cai and R.~H.~Brandenberger,
  %``Evolution of cosmological perturbations and the production of non-Gaussianities through a nonsingular bounce: Indications for a no-go theorem in single field matter bounce cosmologies,''
  Phys.\ Rev.\ D {\bf 92}, no. 6, 063532 (2015)
  %doi:10.1103/PhysRevD.92.063532
  [arXiv:1508.04141 [hep-th]].
  %%CITATION = doi:10.1103/PhysRevD.92.063532;%%

\bibitem{Battarra:2014tga}
  L.~Battarra, M.~Koehn, J.~L.~Lehners and B.~A.~Ovrut,
  %``Cosmological Perturbations Through a Non-Singular Ghost-Condensate/Galileon Bounce,''
  JCAP {\bf 1407}, 007 (2014)
  %doi:10.1088/1475-7516/2014/07/007
  [arXiv:1404.5067 [hep-th]].

%\cite{Liu:2013kea}
\bibitem{Liu:2013kea}
  Z.~G.~Liu, Z.~K.~Guo and Y.~S.~Piao,
  %``Obtaining the CMB anomalies with a bounce from the contracting phase to inflation,''
  Phys.\ Rev.\ D {\bf 88}, 063539 (2013)
  %doi:10.1103/PhysRevD.88.063539
  [arXiv:1304.6527 [astro-ph.CO]].
  %%CITATION = doi:10.1103/PhysRevD.88.063539;%%
  %74 citations counted in INSPIRE as of 22 Jul 2016

%\cite{Liu:2013iha}
\bibitem{Liu:2013iha}
  Z.~G.~Liu, Z.~K.~Guo and Y.~S.~Piao,
  %``CMB anomalies from an inflationary model in string theory,''
  Eur.\ Phys.\ J.\ C {\bf 74}, no. 8, 3006 (2014)
  %doi:10.1140/epjc/s10052-014-3006-0
  [arXiv:1311.1599 [astro-ph.CO]].
  %%CITATION = doi:10.1140/epjc/s10052-014-3006-0;%%
  %30 citations counted in INSPIRE as of 22 Jul 2016

%\cite{Piao:2003zm}
\bibitem{Piao:2003zm}
  Y.~S.~Piao, B.~Feng and X.~m.~Zhang,
  %``Suppressing CMB quadrupole with a bounce from contracting phase to inflation,''
  Phys.\ Rev.\ D {\bf 69}, 103520 (2004)
  %doi:10.1103/PhysRevD.69.103520
  [hep-th/0310206].
  %%CITATION = doi:10.1103/PhysRevD.69.103520;%%
  %110 citations counted in INSPIRE as of 22 Jul 2016

%\cite{Piao:2005ag}
\bibitem{Piao:2005ag}
  Y.~S.~Piao,
  %``A Possible explanation to low CMB quadrupole,''
  Phys.\ Rev.\ D {\bf 71}, 087301 (2005)
  %doi:10.1103/PhysRevD.71.087301
  [astro-ph/0502343].
  %%CITATION = doi:10.1103/PhysRevD.71.087301;%%
  %45 citations counted in INSPIRE as of 22 Jul 2016

  %\cite{Vikman:2004dc}
\bibitem{Vikman:2004dc}
  A.~Vikman,
  %``Can dark energy evolve to the phantom?,''
  Phys.\ Rev.\ D {\bf 71}, 023515 (2005)
  doi:10.1103/PhysRevD.71.023515
  [astro-ph/0407107].
  %%CITATION = doi:10.1103/PhysRevD.71.023515;%%
  %381 citations counted in INSPIRE as of 21 Jul 2016

  %\cite{Easson:2016klq}
\bibitem{Easson:2016klq}
  D.~A.~Easson and A.~Vikman,
  %``The Phantom of the New Oscillatory Cosmological Phase,''
  arXiv:1607.00996 [gr-qc].
  %%CITATION = ARXIV:1607.00996;%%

%\cite{Martin:2003sf}
\bibitem{Martin:2003sf}
  J.~Martin and P.~Peter,
  %``Parametric amplification of metric fluctuations through a bouncing phase,''
  Phys.\ Rev.\ D {\bf 68}, 103517 (2003)
  [hep-th/0307077].
  %%CITATION = HEP-TH/0307077;%%

%\cite{Solomons:2001ef}
\bibitem{Solomons:2001ef}
  D.~M.~Solomons, P.~Dunsby and G.~Ellis,
  %``Bounce behaviour in Kantowski-Sachs and Bianchi cosmologies,''
  Class.\ Quant.\ Grav.\  {\bf 23}, 6585 (2006)
  [gr-qc/0103087].
  %%CITATION = GR-QC/0103087;%%

%\cite{Erickson:2003zm}
\bibitem{Erickson:2003zm}
  J.~K.~Erickson, D.~H.~Wesley, P.~J.~Steinhardt and N.~Turok,
  %``Kasner and mixmaster behavior in universes with equation of state $w \ge 1$,''
  Phys.\ Rev.\ D {\bf 69}, 063514 (2004)
  [hep-th/0312009].
  %%CITATION = HEP-TH/0312009;%%

%\cite{Cai:2013vm}
\bibitem{Cai:2013vm}
  Y.-F.~Cai, R.~Brandenberger and P.~Peter,
  %``Anisotropy in a Nonsingular Bounce,''
  Class.\ Quant.\ Grav.\  {\bf 30}, 075019 (2013)
  [arXiv:1301.4703 [gr-qc]].
  %%CITATION = ARXIV:1301.4703;%%

%\cite{Xue:2010ux}
\bibitem{Xue:2010ux}
  B.~Xue and P.~J.~Steinhardt,
  %``Unstable growth of curvature perturbation in non-singular bouncing cosmologies,''
  Phys.\ Rev.\ Lett.\  {\bf 105}, 261301 (2010)
  [arXiv:1007.2875 [hep-th]].
  %%CITATION = ARXIV:1007.2875;%%

%\cite{Xue:2011nw}
\bibitem{Xue:2011nw}
  B.~Xue and P.~J.~Steinhardt,
  %``Evolution of curvature and anisotropy near a nonsingular bounce,''
  Phys.\ Rev.\ D {\bf 84}, 083520 (2011)
  [arXiv:1106.1416 [hep-th]].
  %%CITATION = ARXIV:1106.1416;%%

%\cite{Koehn:2013upa}
\bibitem{Koehn:2013upa}
  M.~Koehn, J.~L.~Lehners and B.~A.~Ovrut,
  %``Cosmological super-bounce,''
  Phys.\ Rev.\ D {\bf 90}, no. 2, 025005 (2014)
  %doi:10.1103/PhysRevD.90.025005
  [arXiv:1310.7577 [hep-th]].
  %%CITATION = doi:10.1103/PhysRevD.90.025005;%%

%\cite{WilsonEwing:2012pu}
\bibitem{WilsonEwing:2012pu}
  E.~Wilson-Ewing,
  %``The Matter Bounce Scenario in Loop Quantum Cosmology,''
  JCAP {\bf 1303}, 026 (2013)
  [arXiv:1211.6269 [gr-qc]].
  %%CITATION = ARXIV:1211.6269;%%

%\cite{Wilson-Ewing:2013bla}
\bibitem{Wilson-Ewing:2013bla}
  E.~Wilson-Ewing,
  %``Ekpyrotic loop quantum cosmology,''
  JCAP {\bf 1308}, 015 (2013)
  [arXiv:1306.6582 [gr-qc]].
  %%CITATION = ARXIV:1306.6582;%%

%\cite{Amoros:2014tha}
\bibitem{Amoros:2014tha}
  J.~Amor¨®s, J.~de Haro and S.~D.~Odintsov,
  %``$R+\alpha R^2$ Loop Quantum Cosmology,''
  Phys.\ Rev.\ D {\bf 89}, no. 10, 104010 (2014)
  %doi:10.1103/PhysRevD.89.104010
  [arXiv:1402.3071 [gr-qc]].
  %%CITATION = doi:10.1103/PhysRevD.89.104010;%%

%\cite{Sriramkumar:2015yza}
\bibitem{Sriramkumar:2015yza}
  L.~Sriramkumar, K.~Atmjeet and R.~K.~Jain,
  %``Generation of scale invariant magnetic fields in bouncing universes,''
  JCAP {\bf 1509}, no. 09, 010 (2015)
  %doi:10.1088/1475-7516/2015/09/010
  [arXiv:1504.06853 [astro-ph.CO]].
  %%CITATION = doi:10.1088/1475-7516/2015/09/010;%%
  %6 citations counted in INSPIRE as of 02 Aug 2016

%\cite{Chowdhury:2016aet}
\bibitem{Chowdhury:2016aet}
{
  D.~Chowdhury, L.~Sriramkumar and R.~K.~Jain,
  %``Duality and scale invariant magnetic fields from bouncing universes,''
  arXiv:1604.02143 [gr-qc].
  %%CITATION = ARXIV:1604.02143;%%
  %3 citations counted in INSPIRE as of 24 Sep 2016
}

%\cite{Campanelli:2015lqz}
\bibitem{Campanelli:2015lqz}
  L.~Campanelli,
  %``Superhorizon magnetic fields,''
  Phys.\ Rev.\ D {\bf 93}, no. 6, 063501 (2016)
  doi:10.1103/PhysRevD.93.063501
  [arXiv:1512.08600 [astro-ph.CO]].
  %%CITATION = doi:10.1103/PhysRevD.93.063501;%%
  %2 citations counted in INSPIRE as of 02 Aug 2016

\end{thebibliography}
\end{document}